\documentclass[intlimits,twoside,a4paper]{article}

\usepackage[cp1251]{inputenc}

\usepackage[eqsecnum]{cmpj3}

\issue{2022}{25}{4}{42201}
\doinumber{10.5488/CMP.25.42201}

\title[New trends in the nanophysics]%
{New trends in the nanophysics of ferroics, relaxors and multiferroics}
\author[M. D. Glinchuk, L. P. Yurchenko, E. A. Eliseev]{M. D. Glinchuk\orcid{0000-0002-5317-3082}\thanks{Corresponding author: \email{glin@ipms.kiev.ua}.}, L. P. Yurchenko\orcid{0000-0002-0345-9971}, E. A. Eliseev\orcid{0000-0001-8124-8857}}
\address{Institute for Problems of Materials Science, National Academy of Sciences of Ukraine,  3 Omeljan Pritsak St., 03142 Kyiv, Ukraine}


\date{Received June 30, 2022, in final form October 30, 2022}

\begin{document}

\maketitle

\begin{abstract}
The review covers the theoretical and experimental results obtained in the recent years by the scientists with the help of comprehensive investigation of nanoferroics and multiferroics. The main attention will be paid to spontaneous flexoeffects and reentrant phase in nanoferroics as well as to a recently discovered giant magnetoelectric effect in multiferroics.
%
%
\printkeywords
%
\end{abstract}

\section{Introduction}

The review covers the theoretical and experimental results obtained in the recent years with the help of comprehensive investigation of nanoferroics and multiferroics. The main attention will be paid to spontaneous flexoeffects and reentrant phase in nanoferroics and to a recently discovered giant magnetoelectric effect in multiferroics. Being characteristic of all nanoes, geometrical confinement generates many physical effects, which are absent in the bulk ferroic samples. These phenomena are scrupulously described in Chapter 4 of reference~\cite{gli13}. In particular, the authors of~\cite{gli13} considered the appearance of ferromagnetism at room temperature in nanoparticles and thin films of undoped CeO$_2$, HfO$_2$, SnO$_2$, Al$_2$O$_3$ and other nonmagnetic in corresponding bulk oxides. Keeping in mind the detailed description of this phenomenon in Chapter 4 of reference~\cite{gli13} and references given there, we excluded this problem from our review. It appeared that in the recent years, different kinds of  flexo-coupling (flexoelectric, flexomagnetic, flexoelastic) in ferroic nanosamples have been the most popular due to strong geometrical confinement and the influence of the defects.

The flexoelectric effect existing for any solid bodies is known to be the most studied. It was predicted by Mashkevich and Tolpygo~\cite{mas57}. Later on, theoretical study of the flexoelectric effect in bulk crystals was performed by Kogan and Tagantsev~\cite{kog64,tag86,tag91}, experimental measurements of flexoelectric tensor components were carried out by Ma and Cross~\cite{ma01,ma02,ma03} and Zubko et al.~\cite{zub07}. Renovation of the theoretical description for the flexoelectric response of different nanostructures starts from the papers by Catalan and co-workers~\cite{cat04,cat05}, while recent achievements are presented in the papers by Majdoub et al.~\cite{maj08}, Kalinin and Meunier~\cite{kal08} and Lee et al.~\cite{lee11}. In the latter paper, a giant flexoelectric effect (6--7 orders of magnitude larger than the typical value reported for bulk oxides) was discovered in ferroelectric HoMnO$_3$ epitaxial thin film on Pt/Al$_2$O$_3$ substrate. In these papers the flexoelectric effect was considered as a coupling between intrinsic polar properties (e.g., polarization) and the extrinsic factors like the misfit strain relaxation~\cite{cat04,cat05} or the system bending by external forces~\cite{maj08,kal08}. The influence of flexoelectric coupling on the properties of low-dimensional transition metal dichalcogenides was studied by Morozovska et al.~\cite{mor21}. Recent trends and achievements in flexoelectricity research and applications could be found in~\cite{ari22} and in references therein. The coupling between intrinsic parameters, namely, spontaneous polarization gradient inherent to nanosystems and strain, was considered in~\cite{eli09}. The crucial role of the surface in all physical properties of nanosystems including the strong order parameter gradients in ferroic nanostructures~\cite{gli08} inevitably leads to the noticeable influence of  flexocoupling, almost negligible for bulk materials, since the order parameters are usually homogeneous in this case.

Flexomagnetic effect is much less studied in comparison with flexoelectric effect, and only a few relevant papers exist~\cite{luk10,lukas10}. Partly this can be related to the fact that since the inversion of time must be included in consideration, there is symmetry restriction for the existence of flexomagnetic effect. In particular, the existence of time inversion or space inversion uncoupled with other symmetry operations excludes the flexomagnetic effect in para- and diamagnetics. The existence of the symmetry operations coupling or the absence of time inversion (e.g., in antiferromagnetics) makes it possible to get flexomagnetic effect. To find out the non-zero flexomagnetic tensor components, one should consider 90 magnetic classes and perform their symmetry consideration similarly to piezomagnetic and magnetoelectric effects (see Section~4.3.7 in the book~\cite{gli13}).

In what follows we will consider the strong influence of flexoelectric effect in nanoferroics, namely the critical size disappearance at size induced phase transitions and reentrant phase occurrence in nanoferroics. Stabilization of ferroelectric phase with nanoparticles sizes decrease as it was observed earlier in tetragonal BaTiO$_3$ nanospheres of radii 5--50~nm stayed unexplained until the appearance of our paper~\cite{mor16}. Our calculations have shown that the physical mechanism of this exciting phenomenon can be the flexo-chemo effect, being a synergy of the flexoelectric stresses and the chemical pressure induced by ion vacancies via Vegard effect. The coexistence of ferroelectric and relaxor phases due to oxygen vacancies is considered in~\cite{gli18}. The results of our calculations and comparison of the developed theory with experiment can be found in sections 2, 3 and 4; a giant magnetoelectric effect in multiferroics is presented in section 4.2 and was published in our papers~\cite{gli20,lag20}.

\section{Spontaneous flexoeffect}

The most general definition of the direct flexoeffect is the appearance of either polarization~$P$ or magnetization $M$ in response to inhomogeneous mechanical impact, i.e., strain gradient $\partial {u _{ij}}/\partial {x_l}$. The converse flexoeffect corresponds to the appearance of mecha\-nical strain in response to the gradient of either polarization or magnetization, respectively. Therefore, the form of flexoelectric and flexomagnetic effects can be written as:
\begin{equation}
\label{eta}
{\eta _i} = {f_{ijkl}}\frac{{\partial {u_{ij}}}}{{\partial {x_l}}},
\end{equation}
\begin{equation}
\label{uij}
{u_{ij}} = {f'_{ijkl}}\frac{{\partial {\eta _k}}}{{\partial {x_l}}}.
\end{equation}
Here, $\eta = P$ and $M$ stand for flexoelectric and flexomagnetic effects, respectively.

For flexoelastic direct and converse effects ${\sigma _{im}} = {f_{imjkl}}\left( {\partial {u_{jk}}/\partial {x_l}} \right)$ and ${u_{jk}} = {f'_{imjkl}}\left( {\partial {\sigma _{im}}/\partial {x_l}} \right)$, respectively, so that they are defined by 5th rank tensors, while flexoelectric and flexomagnetic effects are defined by 4th rank tensors. In what follows we will pay attention mainly to flexoelectric and flexomagnetic effects in nanostructures. Let us underline that flexocoupling affects both the system response to the external impact and the intrinsic gradient of the order parameters.

Let us note that spontaneous flexocoupling was introduced in~\cite{eli09}, physical mechanism of reentrant phase appearance was considered in~\cite{mor16}, the appearance of morphotropic phase in the relaxor due to oxygen vacancies influence~\cite{gli18}, mechanism of giant magnetoelectric coupling appearance as well as the materials of this phenomenon observation~\cite{gli20,lag20} were proposed by the authors of this review.

\subsection{Analytical theory and comparison of the theory with experiment}

Landau--Ginzburg--Devonshire (LGD) functional bulk (b) and surface (s) densities have a relatively simple form for a nanoparticle with uniaxial ferroelectric polarization $\vec P = \left( 0,0,P_3 \right)$~\cite{eli09}:

\begin{align}
\label{phib}
\Phi _b = \int\limits_V {{G} {\rd^3}r}, 
\end{align}
\begin{align}
\label{phii}
{G} =\, &{\alpha _b}(T)\frac{{P_3^2}}{2} + \beta \frac{{P_3^4}}{4} + \gamma \frac{{P_3^6}}{6} + \frac{{{g_{ij33}}}}{2}\left( {\frac{{\partial \,{P_3}}}{{\partial {x_i}}}\frac{{\partial {P_3}}}{{\partial {x_j}}}} \right) - \frac{{{F_{3ijk}}}}{2}\left( {{P_3}\frac{{\partial \sigma _{ij}^{}}}{{\partial {x_k}}} - \sigma _{ij}^{}\frac{{\partial {P_3}}}{{\partial {x_k}}}} \right)  
\nonumber\\
&-{P_3}\left( {\frac{{E_3^d}}{2} + E} \right) - {Q_{ij33}}{\sigma _{ij}}P_3^2 - \frac{{{s_{ijkl}}}}{2}\sigma _{ij}^{}\sigma _{kl}^{} - {W_{ij}}{\sigma _{ij}}\delta N,
\end{align}
\begin{align}
\label{phis}
{\Phi _S} = \int\limits_S {{\rd^2}r} \left( {\frac{{\alpha ^S}}{2}P_3^2 + \frac{{\beta ^S}}{4}P_3^4 + \mu _{\alpha \beta }^Su_{\alpha \beta } + ...} \right).
\end{align}
The coefficient $\alpha _b(T)$ typically depends on the temperature $T$. Here, we assume the linear dependence, $\alpha _b(T) = \alpha _T(T - T_{\rm C})$, where $T_{\rm C}$ is a Curie temperature. Coefficient $\beta$ sign depends on the ferroelectric transition order, $\gamma > 0$. $E_3^d$ is depolarization field, $Q_{ijkl}$ are the bulk electrostriction tensor coefficients, $g_{ijkl}$ is the gradient coefficients tensor, $F_{ijkl}$ is the flexoelectric strain tensor, $\sigma _{ij}$ is the stress tensor, $W _{ij}$~is the elastic dipole (or Vegard strain) tensor, that is regarded diagonal hereinafter, i.e., $W_{ij} = W\delta _{ij}$ ($\delta _{ij}$~is delta Kroneker symbol). $\delta N = N\left( {\vec r} \right) - N_e$ is the difference between the concentration of defects $N(r)$ in the point $r$ and their equilibrium (average) concentration $N_e$. Surface energy coefficients $\alpha ^S$ and $\beta ^S$ are supposed to be positive and temperature independent, $\mu _{\alpha,\beta}^S$ is the surface stress tensor~\cite{mas57,mar80}, $u_{ij}$ is the strain tensor.

\begin{figure}[htb]
\centerline{\includegraphics[width=0.72\textwidth]{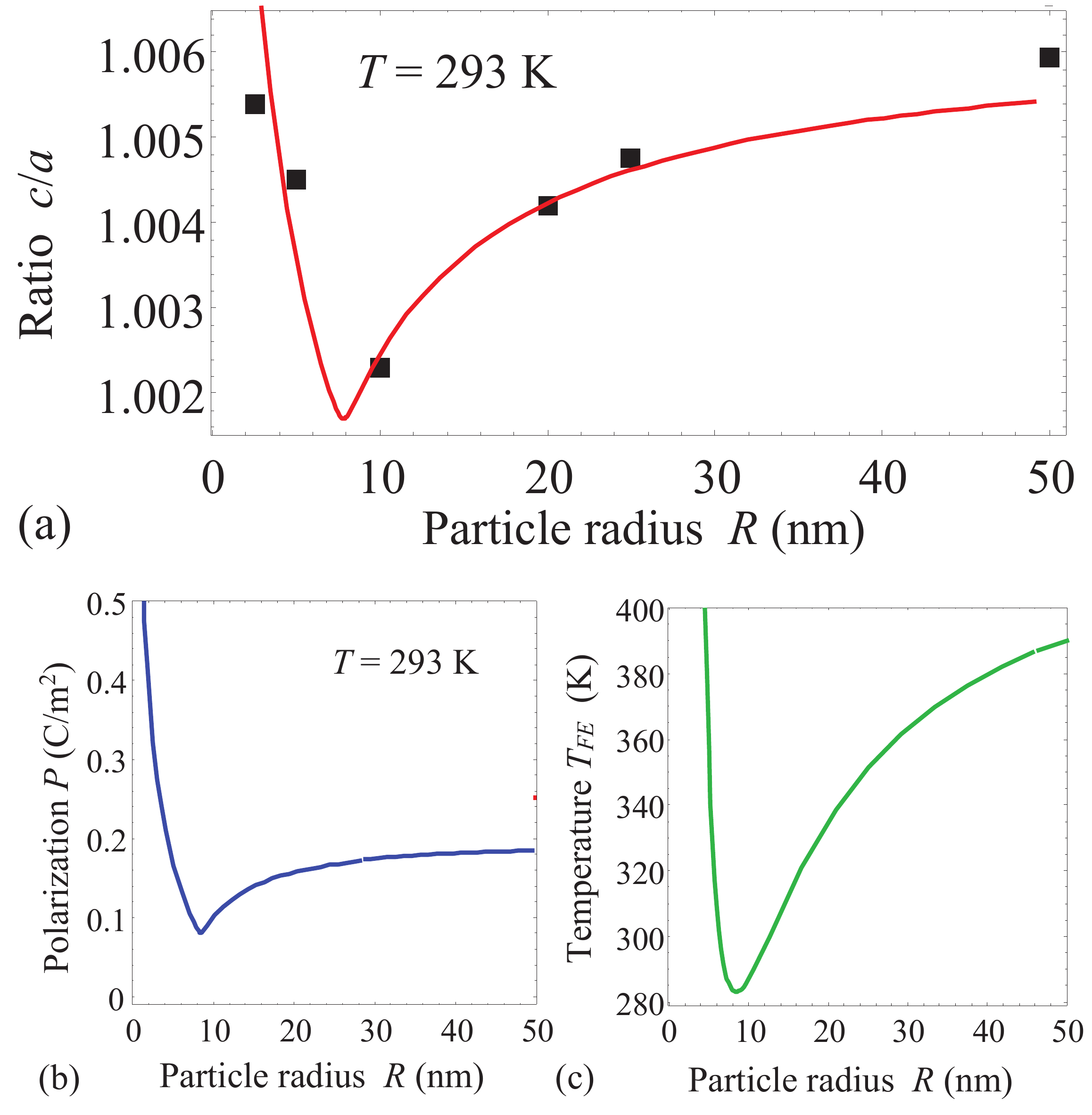}}
\caption{(Colour online) Room temperature tetragonality ratio $c/a$ (a), spontaneous polarization (b), and transition temperature (c) vs. the particle radius $R$. Plot (a) shows the best fit of our theory (solid curve) to experimental data~\cite{zhu12} (symbols). Temperature $T = 293$~K. Adapted from~\cite{mor16}.} \label{fig-smp1}
\end{figure}

Figure~\ref{fig-smp1}~(a) shows the best fit of our theoretical results (solid curve) obtained with the help of equations~(\ref{phib}), (\ref{phis}), which allowed to fit all physical properties with two fitting parameters (see~\cite{mor16} for details) to experimental results on tetragonality~\cite{zhu12} (symbols with error bars). The latter is obtained from X-ray diffraction data on the size and temperature dependence of the lattice constants. Our aim was not to fit well the tetragonality for the smallest particle of radius 2.5~nm, because here the phenomenological continuous approach may be invalid.

Figures~\ref{fig-smp1}~(b) and~(c) illustrate the dependences of the spontaneous polarization and transition temperature on the spherical particle radius calculated for the same fitting parameters as in the figure~\ref{fig-smp1}~(a). Hence, we ``reconstruct'' the polarization and transition temperature from to the best fit of the tetragonality measured experimentally~\cite{zhu12}. The reconstructed dependences clearly demonstrate a strong (more than~2 times) enhancement of polarization and transition temperature for the particle radius less than 10~nm. The reentrant phase region appears for radii $R < 10$~nm. One can see that theoretical figure~\ref{fig-smp1}~(a) rather well describes the spontaneous polarization and transition temperature reconstructed from experimental $c/a$ value.

It is important to underline that the obtained results are very important for both fundamental physics and applications in modern electronic technique.

\section{Relaxor ferroelectrics with perovskite structure}

In this section we consider relaxor ferroelectrics with perovskite structure having a broad application in modern electronic devices~\cite{cros96,park97}. More than 15 years ago the authors of~\cite{deng05} observed the appearance of ferroelectricity in relaxors lead zinc niobate - lead lanthanum zirconium titanate (PZN--PLZT) sintered in nitrogen atmosphere, which induced high concentration of oxygen vacancies.

The recent paper~\cite{gli18} based on the phenomenological theory approach showed that since the oxygen vacancies are known to be elastic dipoles, they affect the elastic and electric fields due to Vegard and flexoelectric couplings. We have shown that a negative Curie temperature $T_{\rm C}^*$ of a relaxor is renormalized by the elastic dipoles due to the electrostriction coupling and could become positive at some large enough concentration of the vacancies. A positive renormalized temperature $T_{\rm C}^R = T_{\rm C}^* + \Delta T$ is characteristic of the ferroelectric state. At $T < T_{\rm C}^R$, all the polar properties could be calculated in the conventional way for ferroelectrics, but the obtained experimental data favor the coexistence of the ferroelectric phase with a relaxor state, i.e., the presence of a morphotropic region in PZN--PLZT relaxor.

Therefore, both the spontaneous flexoelectric effect in nanoferroics appearing due to the surface influence and the flexoelectric effect induced by the elastic field of the defects can be the source of new physical properties important for fundamental physics and for modern technical applications.

Oxygen vacancies in ABO$_3$ ferroelectrics have a great impact on their physical properties and the perovskite structure is capable of conserving the structure stability even for a high concentration of oxygen vacancies. ``B''~cations are usually shifted from central position in the neighborhood of the oxygen vacancy, because in ABO$_3$ structure the size of oxygen ions and so its vacancies used to be larger than the cation ones. To compensate for the loss of the oxygen negative charges, the equivalent amount of B$^{4+}$ cations should be in a B$^{3+}$ state. The PZN--PLZT samples sintered in nitrogen atmosphere appeared to be black and opaque, because off-central Ti$^{4+}$ transforms into color center Ti$^{3+}$. Note that Ti$^{3+}$ can create layers of the ordered dipoles at large concentration of oxygen vacancies. The electrons necessary for Ti$^{4+}$ into Ti$^{3+}$ transformation can be created from ionization of neutral oxygen vacancy ${V_{\rm O}} \to V_{\rm O}^{ \bullet} + e$, $V_{\rm O}^ \bullet  \to V_{\rm O}^{ \bullet  \bullet } + e$, the $V_{\rm O}^ \bullet$ and $V_{\rm O}^{ \bullet  \bullet }$ being positively charged vacancies. Uncharged vacancy $V_{\rm O}$ represents dilatational center which creates a local compressive strain. Since the conductivity is mainly attributed to the electromigration of oxygen vacancies in perovskite ferroelectrics, the measurements of $dc$ conductivity temperature dependence of the NS and OA specimens were carried out in order to estimate the oxygen vacancies concentration and charge states. Since this complex defect can be represented as $V_{\rm O}^{ \bullet \bullet }+2 \textrm{Ti}^{3+}$, it can be observed in high temperature region only. Therefore, we are faced with the existence $V_{\rm O}^{ \bullet \bullet }$ and $V_{\rm O}$ in NS sample with high concentration of oxygen vacancies.

Note that vacancies tend to accumulate in the vicinity of any inhomogeneities, surfaces and interfaces, since the energy of vacancies formation in such places can be much smaller than in the homogeneous volume~\cite{gli13}. In the places of vacancies accumulation they can create sufficiently strong fields, which in turn can lead to the appearance of new phases in relaxors, for example, polar (ferroelectric) ones. On the contrary, in the places where there are few vacancies, the non-polar relaxor remains. Thus, the polar ferroelectric and nonpolar relaxor states coexistence can be realized in this case. 
Let us briefly consider possible mechanisms of ferroelectricity appearance in NS samples of PZN--PLZT. As we discussed above, the oxygen vacancies in this sample are uncharged $V_{\rm O}$, singly and doubly positively charged $V_{\rm O}^ \bullet$ and $V_{\rm O}^{ \bullet \bullet }$, respectively. Because of the necessity of loss oxygen negative charges compensation, approximately an equivalent amount of Ti$^{3+}$ off-central ions is another group of defects. 

If one formally puts charges $+4P\mu$ and $-4P\mu$ into the unoccupied point of Ti$^{4+}$,  the charge $+4e$ makes an ideal lattice and $-4e$ corresponds to the defect in it. A similar formal operation with the addition of $+e$ and $-e$ to Ti$^{3+}$ position  leads to the appearance of an ideal lattice defect $-4P\mu + (\textrm{Ti} ^{3+} + e)$, that is dipole $d_1 = 4es$, where $s$ is Ti$^{3+}$ off-central shift. The existence of electric dipoles will lead to the appearance of ferroelectric phase due to an indirect interaction of dipoles via the soft optic mode, while the soft mode existence in the ferroelectric relaxors will be discussed later.

Keeping in mind that all the electric dipoles in the regions with sizes of the order of correlation radius $r_c$ must be oriented, one can write the criterion of FE phase appearance as $Nr_c^3 \geqslant 1$, where $N$ is concentration of dipoles.

Another possible mechanism of ferroelectricity in the relaxors can originate from inhomogeneous elastic field via flexoelectric effect, namely ${P_i} = {f_{ijkl}}\partial {u_{kj}}/\partial {x_l}$, where $P_i$ is electric polarization component, $\partial {u_{kj}}/\partial {x_l}$ is mechanical strain gradient, $f_{ijkl}$ is the tensor components of flexoelectric effect. Detailed consideration of this mechanism along with the mechanical strain field originated from the oxygen vacancies (Vegard mechanism) contributions to the appearance of ferroelectricity in the relaxors will be given below.

Gibbs potential density of relaxor ferroelectric materials having some hidden soft phonon polar mode has the following form~\cite{gli18}
\begin{align}
 G = \frac{{{a_{ij}}\left( T \right)}}{2}P_iP_j + \ldots + \frac{{g_{ijkl}}}{2}\frac{{\partial {P_i}}}{{\partial {x_j}}}\frac{{\partial {P_k}}}{{\partial {x_l}}} + \frac{{{F_{ijkl}}}}{2}\left( {\sigma _{kl}\frac{{\partial {P_i}}}{{\partial {x_j}}} - {P_i}\frac{{\partial \sigma _{kl}}}{{\partial {x_j}}}} \right) - Q_{ijkl}\sigma _{ij}{P_k}{P_l} 
\nonumber\\
 - \frac{{s_{ijkl}}}{2}\sigma _{ij}\sigma _{kl} - {P_i}E_i\left( {\bf{r}} \right) - u_{ij}^W\left[ {\delta {N_d}\left( {\bf{r}} \right)} \right]{\sigma _{ij}} + {k_{\rm B}}TS\left( {N_d,N_d^ + } \right), 
\label{gi}
\end{align}
where $P_i$ are the components of polarization vector ($i =$1, 2, 3) and $\sigma_{ij}$ is the elastic stress tensor. The summation is performed over all repeated indices. Dielectric stiffness expansion coefficients $a_{ij} (T)$ are positive, because the intrinsic ferroelectricity is absent, but it depends on temperature reflecting the fact that the hidden phonon mode could soften at negative absolute temperatures. This statement follows from the above mentioned fact, that in PZN, a soft mode was observed at $T = 20$~K, so that its frequency could be zero at a negative temperature. Note that extrapolation of PMN soft mode frequency to zero leads to $T_{\rm C} \approx -150$~K. 

Matrix of the gradient coefficients $g_{ijkl}$ is positively defined. $Q_{ijkl}$ is the electrostriction tensor, $s_{ijkl}$ is the elastic compliances tensor, $F_{ijkl}$ is the forth-rank tensor of flexoelectric coupling. In equation~(\ref{gi}), $E_i (r)$ denotes the random electric field. The configuration entropy function $S(x,y)$ is taken as $S\left( {x,y} \right) = y\ln \left( {y/x} \right) - y$ in the Boltzmann--Planck--Nernst approximation; $k_{\rm B} = 1.3807 \times 10^{-23}$~J/K, where $T$ is the absolute temperature. Equations of state $\partial G/\partial {\sigma _{ij}} = - {u_{ij}}$ determine the strains $u_{ij}$. Euler--Lagrange equations $\partial G/\partial {\sigma _{ij}} = 0$ determine the polarization components.

\begin{figure}[htb]
\centerline{\includegraphics[width=0.45\textwidth]{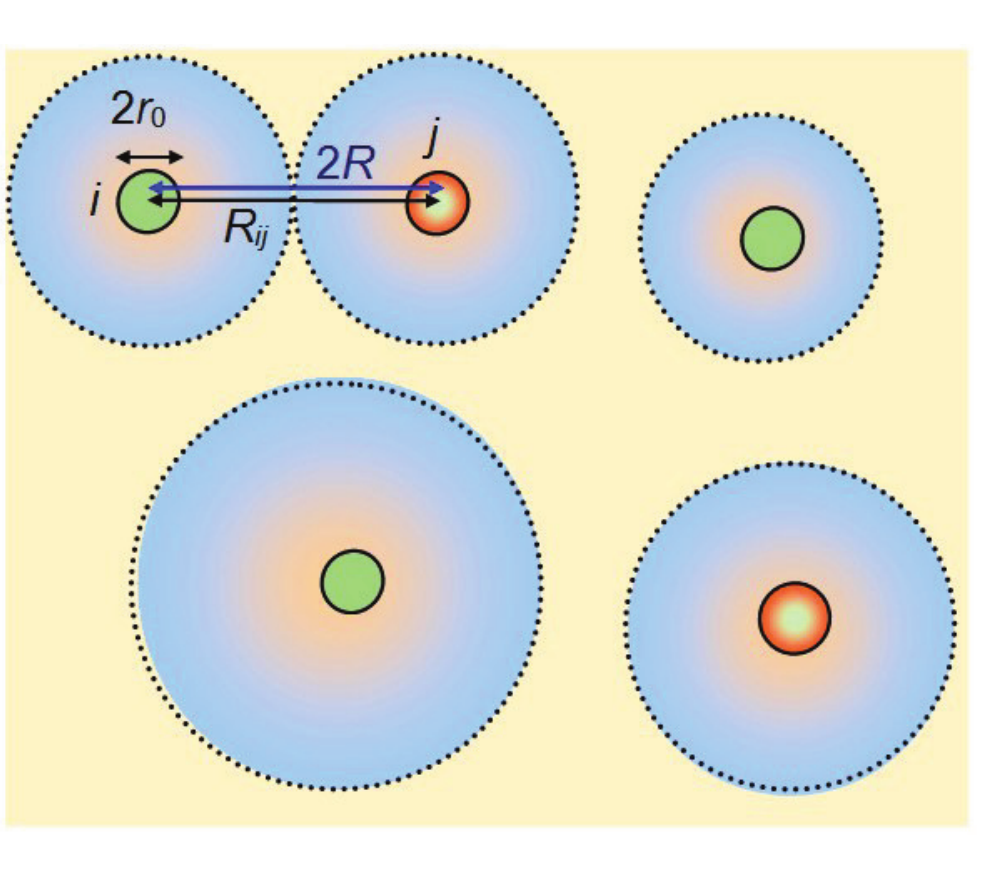}}
\caption{(Colour online) Schematics of the spherical defects with radius $r_0$ embedded into the matrix. The distance between defects ``$i$'' and ``$j$'' is $R_{ij}$. The average distance between defects is $2R$. The average volume per one defect is $V = ({{4\piup }}/{3}){R^3}$. Adapted from~\cite{gli18}.} \label{fig-smp2}
\end{figure}

Equation~(\ref{gi}) includes Vegard-type concentration-deformation energy, $u_{ij}^W\left[ {\delta {N_d}\left( r \right)} \right]{\sigma _{ij}}$, which is determined by the random defects with concentration of $\delta {N_d}\left( r \right) \sim \sum_k \delta \left( {r - {r_k}} \right) - {\bar N_d}$ (e.g., charged or electroneutral oxygen vacancies). The equilibrium concentration of defects is ${\bar N_d} \ll 2.25 \times {10^{28}}$~m$^{-3}$. The average distance between the defect centres $2R$ should be associated with the average volume per inclusion and so it is defined from the relation $({{4\piup }}/{3}){R^3} = 1/{\bar N_d}$. The defect size $r_0$ is much smaller than the average distance $R$, e.g., $r_0$ is ionic radius $\sim (0.1 - 1)$~\AA$^3$ (see figure~\ref{fig-smp2}).

The results~\cite{gli18} showed, that at some concentration of oxygen vacancies, their contribution can be larger than the negative value of temperature $T_{\rm C}^*$ of a relaxor, and so we obtained a positive transition temperature characteristic of a ferroelectric. To transform the relaxors into ferroelectrics, one needs a large enough concentration of oxygen vacancies and Vegard tensor amplitude. Unfortunately, the exact value of negative Curie temperature $T_{\rm C}^*$ is not known and we should discuss some estimations only. It is obvious that even a large enough value of $T_{\rm C}^*$ can be overcome by special choice of oxygen vacancies concentrations and parameters. In such a case, at $T < T_{\rm C}^R$ (ferroelectric phase), all the properties can be written by conventional way on the base of free energy $\Phi = \frac{1}{2}\alpha (T - T_{\rm C}^R){P^2} + \frac{1}{4}\gamma {P^4}$, so that, e. g., polarization ${P^2} = \alpha (T_{\rm C}^R - T)/\gamma $, while at $T > T_{\rm C}^R$, $P = 0$, and we again have a relaxor. For this case, we consider local polarization and electric field induced by Vegard and flexoelectric effects (flexo-chemical coupling). The dependence of the mean square variation of polarization and electric field on concentration of oxygen vacancies showed that flexo-chemical coupling essentially contributes to local polarization and internal electric field.

Keeping in mind the accumulation of oxygen vacancies in the vicinity of different inhomogeneities, we came to the conclusion regarding the oxygen vacancies concentration inhomogeneity. In such a case one can expect coexistence of relaxor state and ferroelectricity Therefore, at some concentration of oxygen vacancies and $T < T_{\rm C}^R$, we are faced with morphotrophic region in PZN--PLZT. 

To resume, the transition to a ferroelectric phase can be induced in a relaxor by the influence of oxygen vacancies being elastic dipoles due to the joint action of electrostrictive and Vegard couplings at some large enough concentration of the vacancies. In the regions where the concentration of vacancies is low, the local polarization and electric field could be induced by the flexo-chemical coupling in dependence on the concentration of oxygen vacancies. Because of inhomogeneity of the vacancies concentration, the coexistence of ferroelectricity and relaxor state can be expected.

\section{Giant magnetoelectric coupling at room temperature in multiferroics}

Nowadays, magnetoelectric coupling is a very important problem for scientists and engineers (see e.g., Feibig et al.~\cite{fei05}). That is why we decided to give a separate introduction and a list of references for the convenience of readers. Moreover, this part is divided into two parts, namely subsection 4.1 based on reference~\cite{gli20} and subsection 4.2 based on reference~\cite{lag20}. The subsections 4.1 and 4.2 differ from one another by the choice of the materials, although the preparation of samples, structural characterization and experimental methods will be the same. The concluding remarks were given as a separate section.

\subsection{Room-temperature ferroelectricity, superparamagnetism and large magnetoelectricity in solid solution PbFe$_{1/2}$Ta$_{1/2}$O$_3$ with (PbMg$_{1/3}$Nb$_{2/3}$O$_3$)$_{0.7}$(PbTiO$_3$)$_{0.3}$}

A strong magnetoelectric (ME) coupling existing at room temperature is especially vital for novel functional device fabrication~\cite{spa08,sco12,che14,sco13,gha21}. A straightforward way of increasing the magnitude of the ME response is to choose the components with large magnetostrictive and piezoelectric coefficients. For a long time, mainly composite multiferroics on the base of PbZr$_{1-x}$Ti$_x$O$_3$ (PZT) or (PbMg$_{1/3}$Nb$_{2/3}$O$_3$)$_{0.7}$(PbTiO$_3$)$_{0.3}$ (PMN--PT) has been used. The latter component produces the largest ME effect and has been often used in the multilayer multiphase (ferroelectric/ferromagnetic) structures. The reason is the large piezoelectricity of PMN--PT in the morphotropic region with the coexistence of the relaxor and ferroelectric phases. For a single crystal, it is 7 times larger than the piezoelectricity of PZT~\cite{fei05}. For ceramic materials, the piezoelectricity in PMN--PT is 2 times larger than that in PZT~\cite{ryu02}.

In the recent years, considerable attention of scientists and engineers was paid to ferroelectric antiferromagnets PbFe$_{1/2}$Ta$_{1/2}$O$_3$ (PFT), $T_N \approx 130$--180~K, and PbFe$_{1/2}$Nb$_{1/2}$O$_3$ (PFN), $T_N \approx 140$~K~\cite{smo58,bok62} and their solid solutions with PbZr$_{0.53}$Ti$_{0.47}$O$_3$~\cite{san11,mar12,eva13,san13,eva14,sch14,sch16,sch17}. Some of these solid solutions exhibit room-temperature multiferroism and large enough ME coupling, which includes a mixture of linear and biquadratic contributions. The ME coupling was theoretically analyzed in references~\cite{gli14,gli16} and was described by second and fourth rank tensors, namely, $\mu _{ij}P_iM_j$ and $\xi _{ijkl} P_i P_j M_k M_l$ ($P$ is polarization and $M$ is magnetization). The authors of papers~\cite{gli14,gli16} have shown that large ME effect and the appearance of magnetization originate from the nanostructure of the considered materials. Another mechanism of the appearance of magnetization in chemically disordered antiferromagnetic multiferroics was considered in works~\cite{kuz14,kuz16}. It was shown that antiferromagnetically interacting Fe$^{3+}$ ions may form superstructures having a ferrimagnetic ground state. Such a structure has a different number of nonequivalent Fe positions in a unit cell. As a result, the ground state magnetization may reach several B per Fe spin. Experimental indications on the ferrimagnetic superstructure formation were reported in reference~\cite{lag14} for PbFe$_{1/2}$Sb$_{1/2}$O$_3$. The F-center exchange mechanism was proposed in
\cite{coey04}. It provides an explanation for the existence of ferromagnetism in oxides due to the presence of oxygen vacancies.

Recently, the attention of scientists and engineers was attracted to the paramagnetoelectric (PME) effect introduced by Hou and Blombergen~\cite{hou65} described by the term $\lambda _{ijk} P_i M_j M_k$. The results of experimental and theoretical studies of this effect in PFN and its solid solution with PbTiO$_3$ were published in references~\cite{lag16,lag17}. 

It was not excluded that the replacement of PZT by PMN--PT in solid solution with PFT could lead to an essential increase of ME effect due to a larger piezoelectric coefficient in PMN--PT. Surprisingly, to the best of our knowledge, there is no information about the synthesis of the (PFT)$_x$(PMN--PT)$_{1-x}$ solid solution. Probably the more complex characteristics of PMN--PT than PZT, and thus the more complicated mechanism of ME effect, were the main reason for the fact. On the other hand, both components of the (PFT)$_x$(PMN--PT)$_{1-x}$ solid solution were broadly studied (see e. g.,~\cite{fal05,cho07,bha12,lam04,zhu00,choi89,feng06}). In particular, it was shown that multiferroic PFT is antiferromagnetic-ferroelectric with ferroelectric phase transition at $T_{\rm C} \approx 250$~K~\cite{ryu02}. The considered PMN--PT is non-magnetic with a maximum of dielectric permittivity at $T_m \approx 410$--420~K~\cite{choi89,ern87}.

The aim of this study is to briefly describe the synthesis methodology of the novel single-phase multiferroic, to characterize the obtained samples and to present the results of experimental and theoretical investigation of its properties, namely, polar, magnetic and magnetoelectric characteristics.

\begin{figure}[htb]
	\centerline{\includegraphics[width=0.95\textwidth]{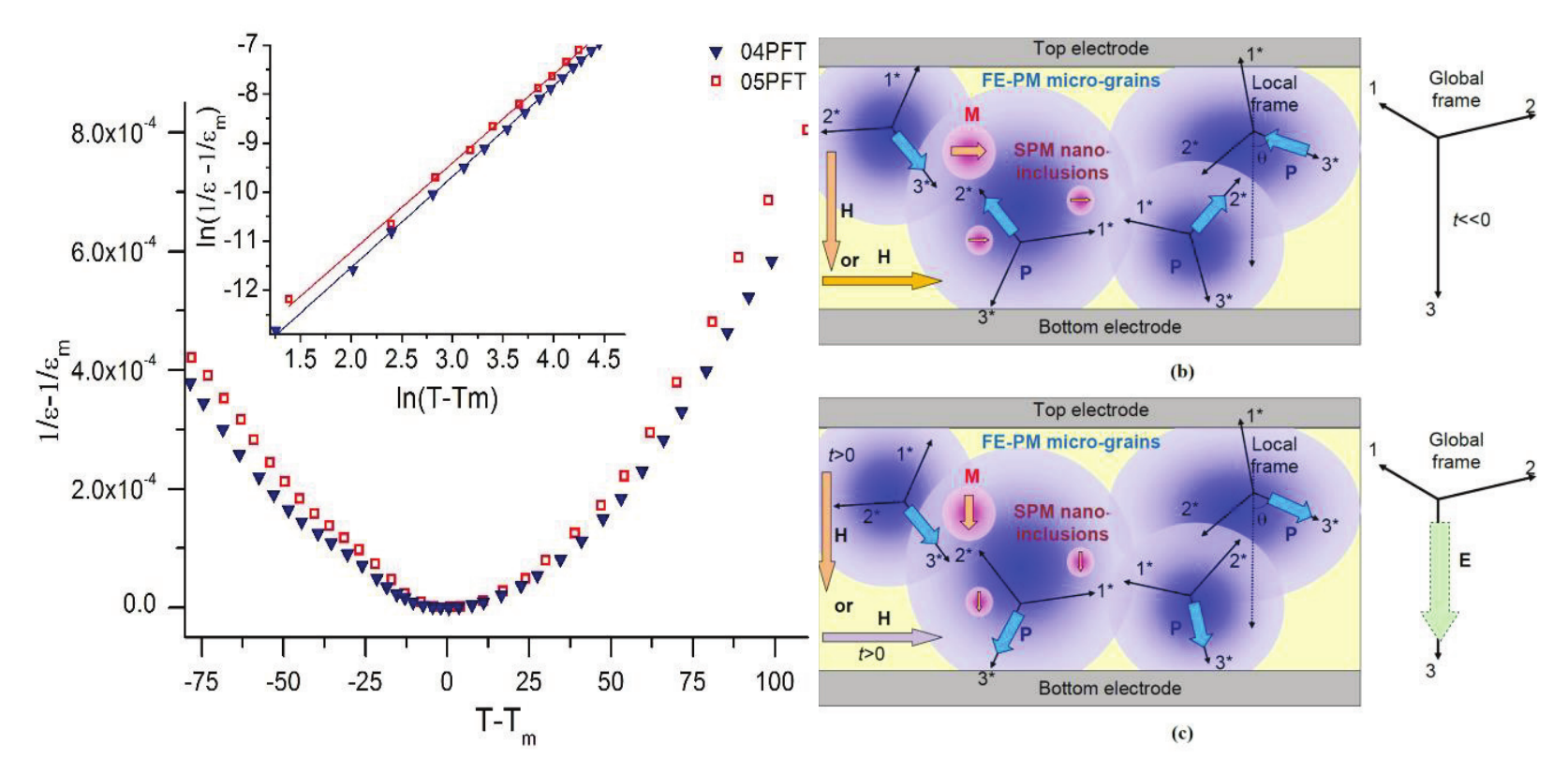}}
	\caption{(Colour online) (a) Temperature dependences of the reciprocal dielectric permittivity of the (PFT)$_x$(PMN--PT)$_{1-x}$ ($x= 0.4$, 0.5) compounds; the inset shows logarithmic dependence of $1/ \varepsilon - 1/ \varepsilon _m$ on $T - T_m$. Adapted from~\cite{gli20}. (b) Pristine nano-grained ceramics before prepoling. The angles $\theta$ between the grain polarization $P$ and global axis $3'$ vary between 0 and 180 degrees. The static magnetic field $H$ was applied to measure $M(H)$. (c) The ceramics after the pre-poling in a strong electric field $E$. The field was applied and then removed. The angle $\theta$ changes between 0 and 90 degrees after strong pre-poling. For $m3m$ parent symmetry, the local axes can be re-numbered in such a way that the axis with minimal angle to the global axis 3 can be labeled as axis 3*. The low-frequency magnetic field $H$ is applied after pre-poling to measure the ME current.} \label{fig-smp3}
\end{figure}

As seen in figure~\ref{fig-smp3}, the temperature dependence of the inverse dielectric permittivity, measured for both samples at 1~kHz does not follow the linear behavior predicted by the Curie--Weiss law for a certain temperature range above $T_m$. In this case, the modified Curie--Weiss law can be used to describe the temperature dependence of the dielectric permittivity~\cite{san01}:
\begin{align}
\label{em}
\frac{1}{\varepsilon } = \frac{1}{{{\varepsilon _m}}} + {\left( {\frac{{T-{T_m}}}{C}} \right)^\gamma },\quad ({\varepsilon _m} = {\varepsilon '_{\max }}).
\end{align}
Parameter $\gamma$ characterizes the degree of phase transition diffuseness, and its values lie in the region from 1 (for ferroelectrics) to 2 (for relaxors). Fitting of the experimental data gives the $\gamma$ values equal to $\sim$1.85 for $x = 0.4$ and $\sim$1.8 for $x = 0.5$. As shown in~\cite{shv06}, the diffuse phase transition could be considered as an intermediate state in which both ferroelectric and relaxor phases can be presented simultaneously. This allows us to suggest that both relaxor and normal ferroelectric phases coexist in the studied compounds, and an increase in the PFT content leads to an increase in the ferroelectric phase contribution.

\begin{figure}[htb]
\centerline{\includegraphics[width=0.68\textwidth]{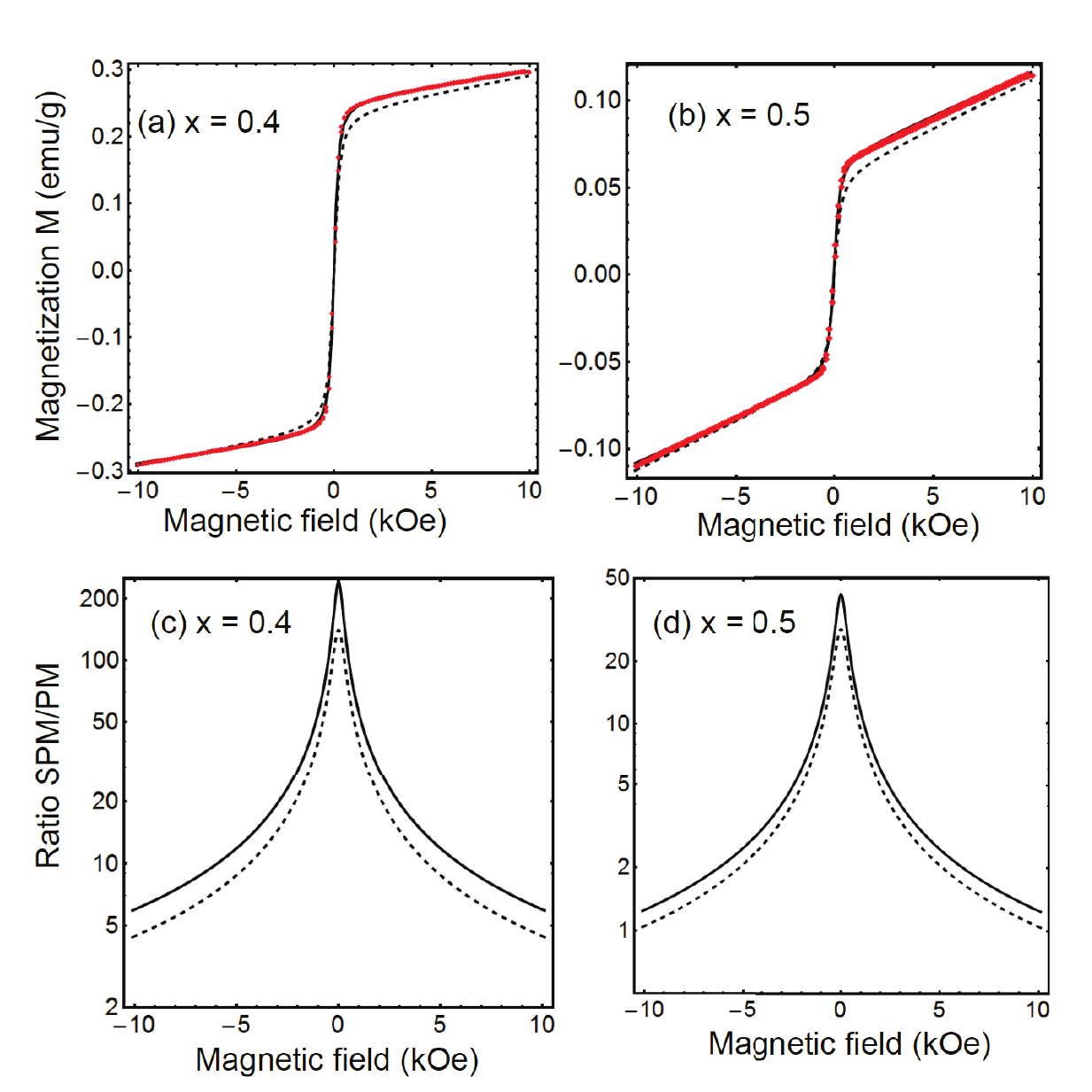}}
\caption{(Colour online) Magnetization dependence on magnetic field for solid solution (PFT)$_x$(PMN--PT)$_{1-x}$ for $x = 0.4$ (a) and $x = 0.5$ (b) at $T = 293$~K. Small red symbols are experimental data~\cite{gli20}, solid and dashed curves are the fitting using Langevin and Brillouin functions, respectively, for describing the superparamagnetic contribution. Plots (c) and (d) show the ratio of the superparamagnetic (SPM) to paramagnetic (PM) contributions to magnetization for $x = 0.4$ and 0.5, respectively.} \label{fig-smp4}
\end{figure}

The magnetic response of solid solutions (PFT)$_x$(PMN--PT)$_{1-x}$ is due to the presence of octahedrally coordinated Fe$^{3+}$ ions having $3d^5$ electronic $d$-shell configuration in $S$-state, spin $S_{\textrm{Fe}} = 5/2$ and $g$-factor $g \approx 2$. Their fraction in the formula unit is $x/2$. Figures~\ref{fig-smp4} (a) and (b) show magnetization isotherms $M(H)$ at $T = 293$~K for the compositions $x = 0.4$ (a) and $x = 0.5$ (b). The $M(H)$ isotherms have a qualitatively similar look for $x = 0.4$ and $x = 0.5$. Symbols are experimental data~\cite{gli20}. For all considered temperatures, the curves are anhysteretic, i.e., reversible. We see that the measured magnetization may be presented as a sum of a paramagnetic contribution that is proportional to the field $M_p(H,T) = \chi _p(T)H$ and of superparamagnetic-like contribution $M_s(H,T)$ that saturates at the field of the order of 1 kOe. The absence of noticeable hysteresis for the observed curves allows us to consider nonparamagnetic contributions as superparamagnetic ones registered at $T \gg T_b$, $T_b$ being a blocking temperature~\cite{bed09,wie03}. Therefore, the blocking temperature is $T_b \ll 293$~K. 

Solid and dashed curves in figures~\ref{fig-smp4} (a) and (b) are theoretical fitting using Langevin and Brillouin functions, respectively, for the description of the superparamagnetic contribution. Herein below we assume the following dependence of the magnetization $M$ on applied quasi-static magnetic field $H$:
\begin{align}
\label{mh}
M\left( H \right) = {M_{PM}}\left( H \right) + {M_{SPM}}\left( H \right) \approx {\chi _m}H + \int\limits_{_{{\mu _{\min }}}}^{_{{\mu _{\max }}}} {F\left( {\mu  - {\mu _S}} \right){M_P}\left( \mu  \right)f\left( {\frac{{\mu H}}{{{k_{\rm B}}T}}} \right)} \rd\mu  + \delta M,
\end{align}
where the first term is a purely paramagnetic (PM) contribution of effective ``host-matrix'', and the second is the superparamagnetic (SPM) contribution of the nanoregions. The value $\delta M$ can be non-zero reflecting a very small device-related systematic error (shift).

\emph{PM contribution}. Note that the linear approximation for PM contribution,
\begin{align}
\label{mpm}
M_{PM}\left( H \right) \approx {\chi _m}H,
\end{align}
is valid for magnetic fields $\left| H \right| \ll ({k_{\rm B}}T)/(g{\mu _B}S)$, where $\mu _B$ is Bohr magneton, $k_{\rm B}$ is Boltzmann constant, and factor $g$ is taken $\approx{2}$. At room temperature, the fields should be much smaller than 800 kOe. The number of Fe$^{3+}$ ions in the PM part of the sample can be obtained from the Curie constant $C_{CW}$ value in the Curie-Weiss law for the ${\chi _m} = {C_{CW}}/(T - \theta )$. The temperature $\theta$ is negative in case of the antiferromagnetic interaction between the ionic spins. Constant $C_{CW} = {N_{PM}}\mu _p^2/3{k_{\rm B}} \approx 4{N_{PM}}\mu _B^2{g^2}S\left( {S + 1} \right)/3{k_{\rm B}}$, where $N_{PM}$ is the number of paramagnetic spin $S$ in a unit mass volume. Hence, $N_{PM} = 3{k_{\rm B}}{C_{CW}}/\left[ {4\mu _B^2{g^2}S\left( {S + 1} \right)} \right]$.

\emph{SPM contribution}. Following Binder and Young~\cite{bin86}, and Wiekhorst et al.~\cite{wie03}, the integration (or averaging) in the last term in equation~(\ref{mh}) reflects the fact that the number of elementary spins, which contribute to the magnetic moment $\mu$ of a given SPM no-regions [and thus to its super-spin] should be different for different inclusions. The magnetic moment can fluctuate around the average value $\mu _S (T)$ in dependence on the sharpness of its distribution function $F \left( \mu - \mu _S \right)$. The function is defined as $\mu _S\left( T \right) = \int_{{\mu _{\min }}}^{{\mu _{\max }}} {F\left( {\mu  - {\mu _S}} \right)\mu \rd\mu }$. The magnetization amplitude $M_P\left( \mu  \right)$ of SPM contribution is equal to $M_P = {N_S}\mu$, where $N_S$ is the average number of super-spins in one gram of the material having the average magnetic moment $\mu _S (T)$.

Depending on a spin, one can use Langevin (LF) or Brillouin (BF) functions for the function $f(x)$ describing SPM contribution:
\begin{align}
\label{funcf}
f\left( x \right) = \left\{ \begin{array}{l}
 \coth \left( x \right) - \frac{1}{x},\quad {\rm{LF}}{\rm{,}} \\ 
 \frac{{2S + 1}}{{2S}}\coth \left( {\frac{{2S + 1}}{{2S}}x} \right) - \frac{1}{{2S}}\coth \left( {{x}/{{2S}}} \right), \quad S \gg 1, \quad {\rm{BF}}.
 \end{array} \right.
\end{align}
As one can see, BF transforms into LF in the limit $S\gg1$, since $\coth \left[ {x(2S + 1)/2S} \right] \approx \coth \left( x \right)$ and $\coth \left( {x/2S} \right) \approx 2S/x$.

In the simplest assumption, when all SPM particles are formed by the same Fe$^{3+}$ ions with $S = 5/2$ and $g = 2$, one can obtain that in average each of these particles contains $n_{\rm{SPM}} = {\mu _S}\left( {T \to 0} \right)/[ {2{\mu _B}\sqrt {S\left( {S + 1} \right)} } ]$ ions and the number of spins (ions) that is included in SPM ensemble is $N_{\rm {SPM}} = n_{\rm {SPM}}{N_S}$. The ratio of the ion Fe$^{3+}$, which belongs to PM and SPM ensembles in the sample, is equal to:
\begin{align}
\label{npmnspm}
\frac{{{N_{{\rm{PM}}}}}}{{{N_{{\rm{SPM}}}}}} = \frac{{2{\mu _B}\sqrt {S\left( {S + 1} \right)} {N_{{\rm{PM}}}}}}{{{N_S}{\mu _S}\left( {T \to 0} \right)}} \equiv \frac{{2{\mu _B}\sqrt {S\left( {S + 1} \right)} }}{{{N_S}{\mu _S}\left( {T \to 0} \right)}} \cdot \frac{{3{k_{\rm B}}{C_{CW}}}}{{4\mu _B^2{g^2}S\left( {S + 1} \right)}} \equiv \frac{{3{k_{\rm B}}{C_{CW}}}}{{2\mu _B^{}{g^2}\sqrt {S\left( {S + 1} \right)} {N_S}{\mu _S}}}.
\end{align}

Figures~\ref{fig-smp4} (c) and (d) show the ratio of SPM to PM contributions to magnetization, i.e., the dimensionless parameter
\begin{align}
\label{alh}
\alpha \left( H \right) = \frac{{{M_P}}}{{\chi _mH}}f\left( {\frac{{{\mu _S}H}}{{{k_{\rm B}}T}}} \right),
\end{align}
for $x = 0.4$ and 0.5, respectively.

\begin{figure}[htb]
\centerline{\includegraphics[width=0.75\textwidth]{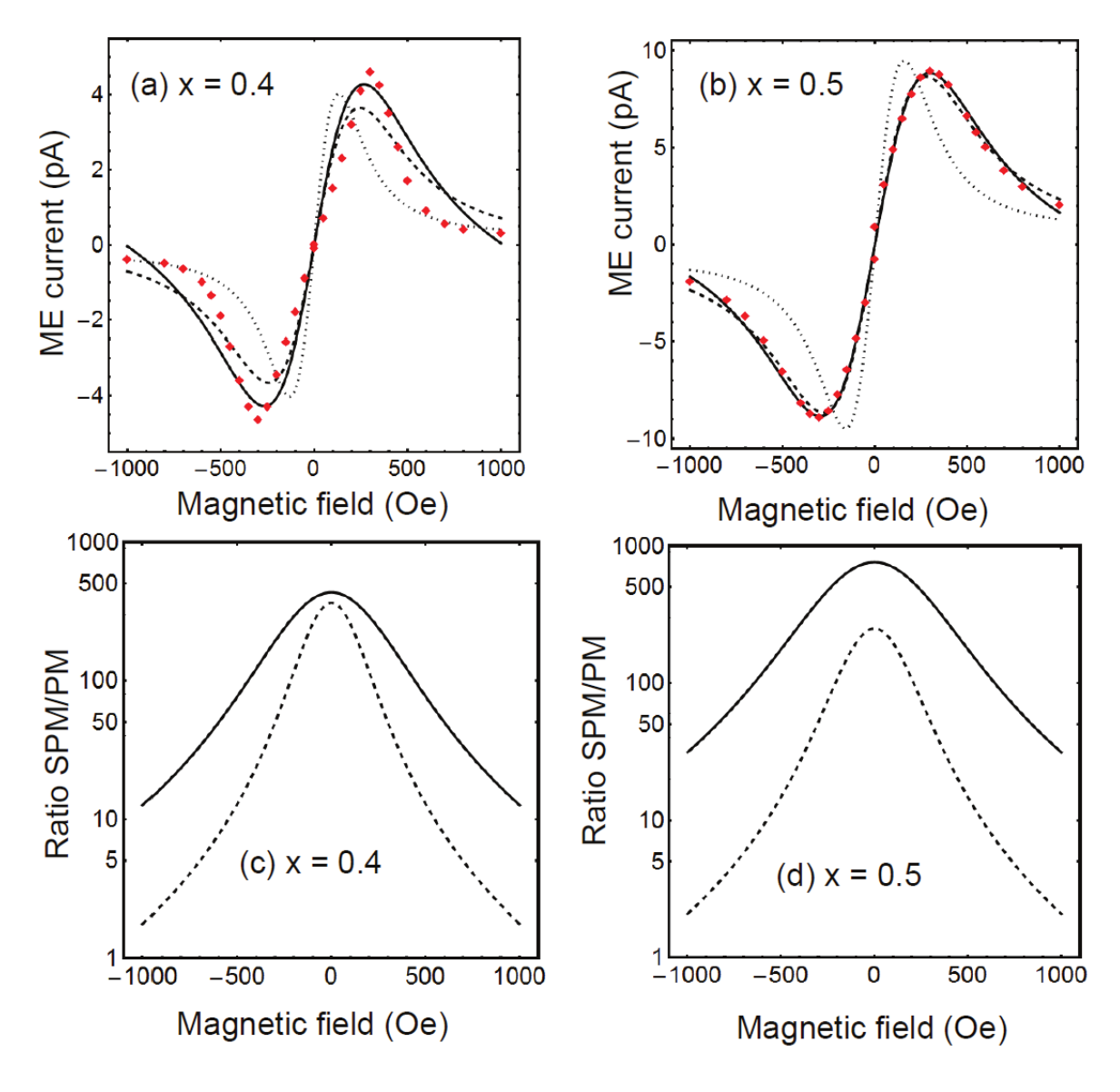}}
\caption{(Colour online) Experimental data~\cite{gli20} (symbols) for ME current in (PFT)$_x$(PMN--PT)$_{1-x}$ ceramics at $T = 293$~K for $x = 0.4$ (a) and $x = 0.5$ (b). Dotted, dashed and solids curves are fitted using~LF. Plots (c) and (d) show the ratio of SPM to PM contributions to ME current calculated for $x = 0.4$ and 0.5, respectively.} \label{fig-smp5}
\end{figure}

Figure~\ref{fig-smp5} shows ME current as a function of the applied $dc$ magnetic field in a PFT--PMN--PT ceramics for the compositions $x = 0.4$ (a) and $x = 0.5$ (b). Symbols are experimental data~\cite{gli20}. One can see that ME signal sharply increases with an increase of the $dc$ magnetic field in the range of $\pm$300 Oe, then it saturates in value and decreases down to almost zero at fields larger than $\pm$3000 Oe. Note, that the ME current does not change too much with temperature lowering from 293~K down to 120~K~\cite{gli20}. Only the current peak position moves from $\pm$300 Oe to $\pm$400 Oe when the temperature changes from~293~K to 120~K. The measurements show that the main contribution to the ME current is caused by the superparamagnetic phase while the contribution of isolated Fe$^{3+}$ spins from the paramagnetic phase is negligibly small due to their much lower magnetic moment. 

Solid and dashed curves in figures~\ref{fig-smp5} (a), (b) are theoretical fitting using Langevin and Brillouin functions. Indeed, using equations~(\ref{mh}) and (\ref{funcf}), one can suppose that the dependence of ME current on the magnetic field $H$ could be described with the following equation
\begin{align}
{I_{ME}} &\approx {I_0}\left\langle {\frac{{\partial M_{\rm{PM}}^2}}{{\partial H}}} \right\rangle  + {I_1}\left\langle {\frac{{\partial M_{\rm{SPM}}^2}}{{\partial H}}} \right\rangle  
 \nonumber\\ 
  &\approx 2{I_0}\chi _m^2H + 2{I_1}\int\limits_{_{{\mu _{\min }}}}^{_{{\mu _{\max }}}} {M_P^2f\left( {\frac{{\mu H}}{{{k_{\rm B}}T}}} \right)\frac{\partial }{{\partial H}}f\left( {\frac{{\mu H}}{{{k_{\rm B}}T}}} \right)F\left( {\mu  - {\mu _S}} \right)\rd\mu }.
\label{ime}
\end{align}
Here, $f(x)$ is a LF or BF given by equations~(\ref{funcf}). Since the phenomena are sample-dependent, the fitting parameters in equations~(\ref{ime}) are $\chi _m$, $M_P$, $\mu _S$, $I_0$, $I_1$, and the distribution function $F\left( \mu  - {\mu _S} \right)$, while the signs of $I_0$ and $I_1$ can be arbitrary. 

Figures~\ref{fig-smp5} (c), (d) show the ratio of SPM to PM contributions to ME current, dimensionless parameter
\begin{align}
\label{gamma}
\gamma \left( H \right) = \frac{{{I_1}M_P^2}}{{{I_0}\chi _m^2H}}f\left( {\frac{{{\mu _S}H}}{{{k_{\rm B}}T}}} \right)\frac{\partial }{{\partial H}}f\left( {\frac{{{\mu _S}H}}{{{k_{\rm B}}T}}} \right),
\end{align}
calculated from equations~(\ref{gamma}) for $x = 0.4$ and 0.5, respectively.

Note that the strong scattering of ME effect values in PFT--PZT solid solutions made it cumbersome to perform a direct comparison with our results obtained for PFT--PMN--PT solid solution. Our study demonstrates that multiferroics with superparamagnetic phase can be considered as promising materials for applications along with composite multiphase (ferroelectric/ferromagnetic) structures. Since the ME response in multiferroics with the superparamagnetic phase is proportional to $\rd M^2/\rd H$, its value can be amplified by many orders due to a sharp change of magnetization with the field. The measurements and theoretical analysis show that the main contribution to the ME current is caused by the superparamagnetic phase.

\subsection{Giant magnetoelectric response in multiferroics with coexistence of superpara\-magnetic and ferroelectric phase at room temperature}

\begin{figure}[htb]
\centerline{\includegraphics[width=0.52\textwidth]{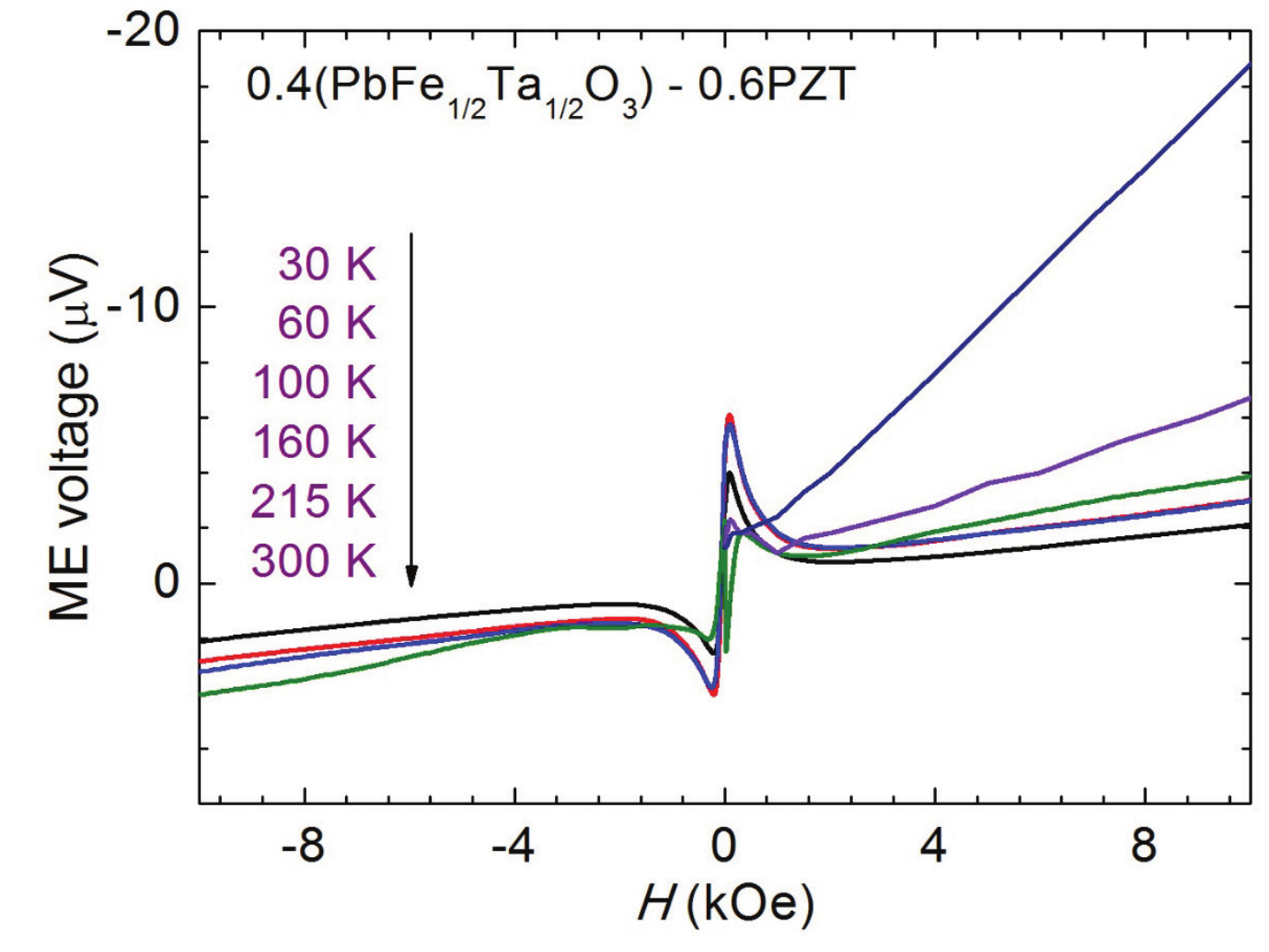}}
\caption{(Colour online) ME voltage as a function of the applied bias magnetic field measured for 0.4PFT--0.6PZT ceramics at temperatures from 300~K down to 6~K. Adapted from~\cite{lag20}.} \label{fig-smp6}
\end{figure}

\begin{figure}[htb]
\centerline{\includegraphics[width=0.53\textwidth]{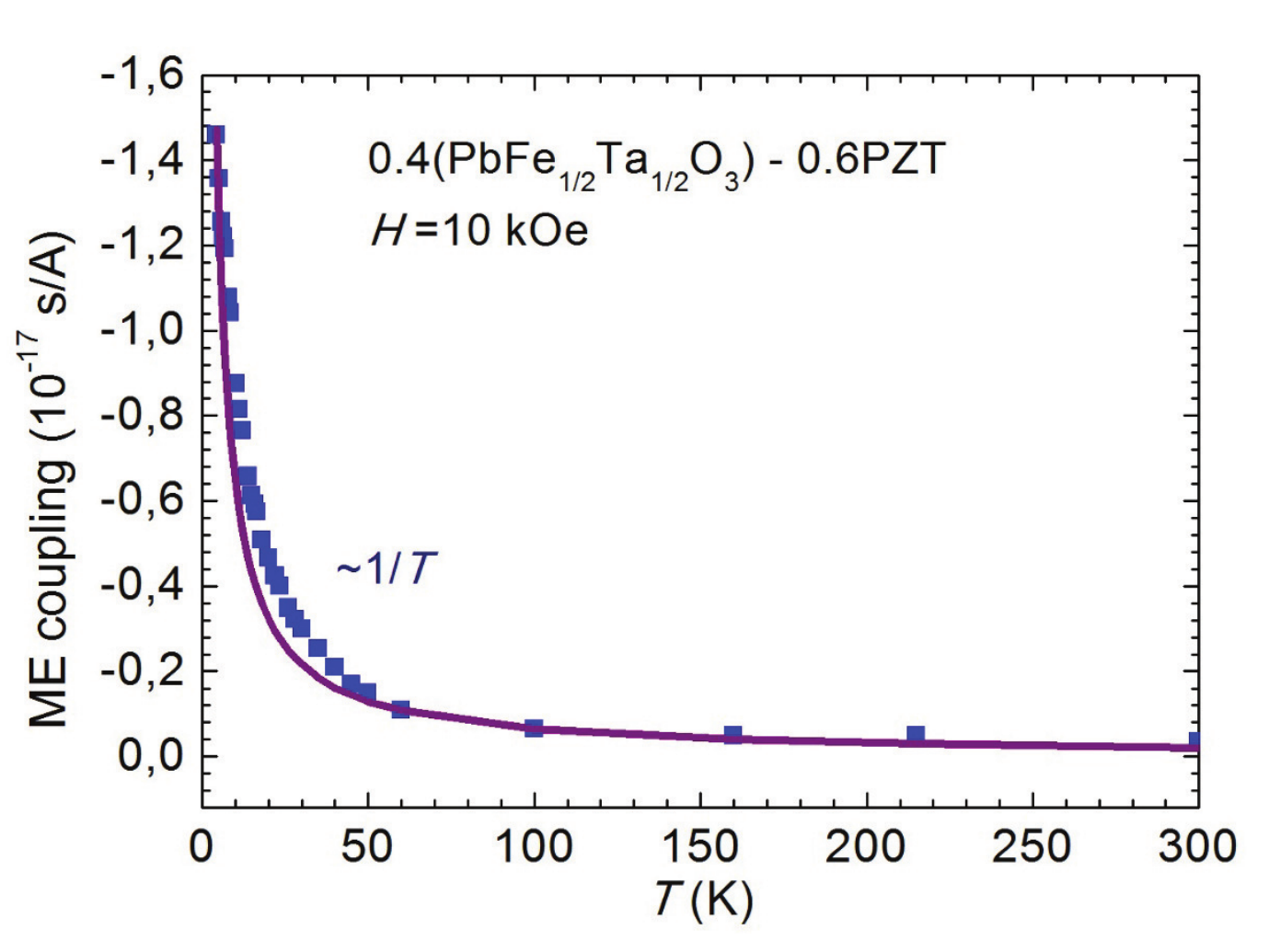}}
\caption{(Colour online) Temperature dependence of the PME coupling coefficient of 0.4PFT--0.6PZT ceramics at the bias magnetic field of 10 kOe. Square symbols and smooth line are measured and calculated data, respectively. Adapted from~\cite{lag20}.} \label{fig-smp7}
\end{figure}

Another example of ME materials where the ME coupling is quite large is the solid solution of PFT with Pb(ZrTi)O$_3$ (PZT). In particular, the composition 0.4PFT--0.6PZT was studied. Figure~\ref{fig-smp6} shows the dependence of the ME voltage as a function of bias magnetic field at temperatures from 300~K down to~6~K. One can see that the behavior of the ME response in the magnetic field at $T > 100$~K is very similar to that of the 0.4PFT--0.6(PMN--PT) ceramics presented in subsection 4.1. Namely, there is a strongly nonlinear ME response in small magnetic fields and the linear part at $H > \pm 1$ kOe. This linear part of the ME signal is related to the paramagnetoelectric contribution in the ME signal. The paramagnetoelectric contribution strongly increases at the temperature decrease. Its temperature dependence (the calculated ME coupling coefficient) measured at the magnetic field of 10 kOe is shown in figure~\ref{fig-smp7}. It follows well the temperature dependence of the magnetic susceptibility (or its square) in the paramagnetic phase created by isolated Fe$^{3+}$ ions. The ME coefficient in this phase tends to infinity as the temperature approaches 0~K due to the relation $\chi \sim 1/T$. However, its actual value is, of course, limited by a saturated magnetization of spins in the magnetic field. 

\begin{figure}[htb]
\centerline{\includegraphics[width=0.87\textwidth]{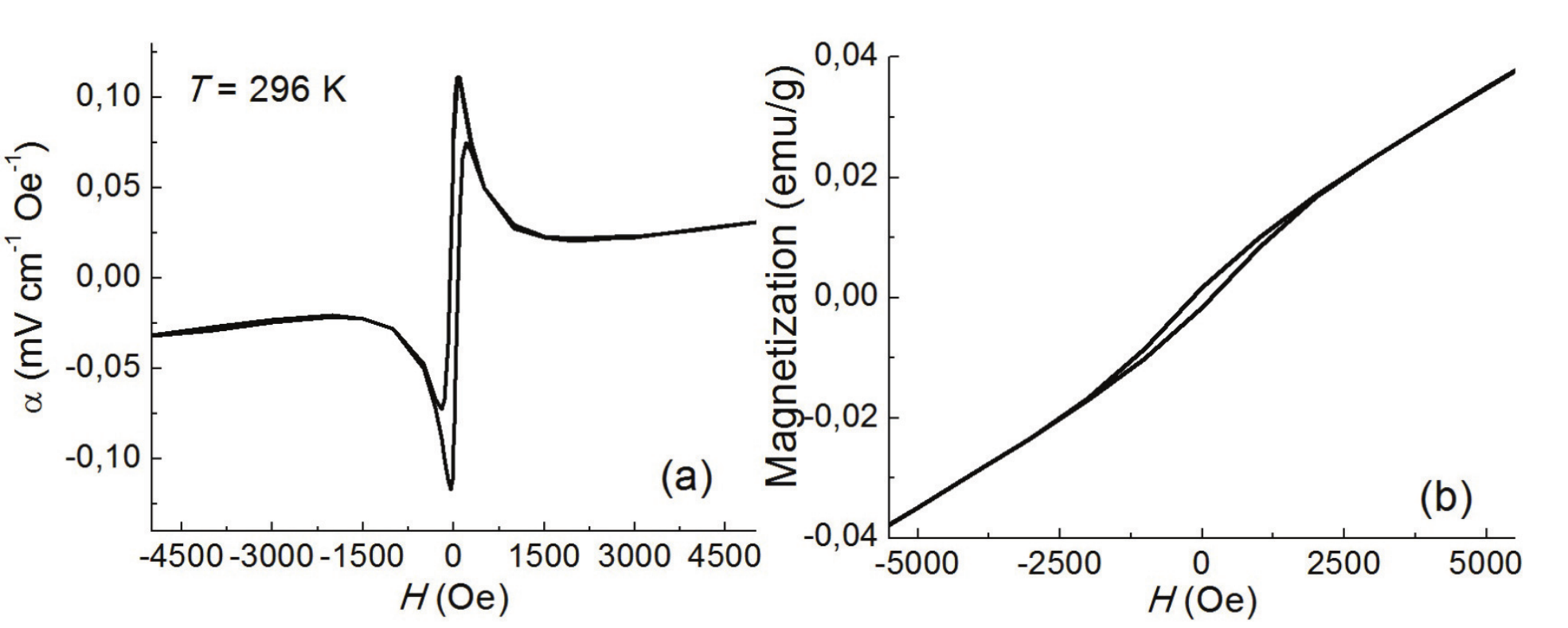}}
\caption{Field dependence of the linear ME coefficient (a) and magnetization (b) measured for 0.4PFT--0.6PZT ceramics at 296~K. Adapted from reference~\cite{lag20}.} \label{fig-smp8}
\end{figure}

Another part of the ME signal is related to superparamagnetism or rather weak magnetism of the 0.4PFT--0.6PZT ceramics. Figure~\ref{fig-smp8} shows this ME signal at small magnetic fields together with magnetization. The magnetization nonlinearly changes at $H < \pm 2$~kOe and shows a hysteresis with slim magnetization loop and remanent magnetization $\approx 0.004$~emu/g. Similar hysteresis is seen in the ME signal. Obviously, the superparamagnetic clusters in 0.4PFT--0.6PZT ceramics are larger in volume than those in 0.4PFT--0.6(PMN--PT) and some interaction may also exist between them. This leads to freezing (thermal blocking) of the superparamagnetic moments of the clusters even at room temperature. Since the remanent magnetization is non-zero, the linear ME effect is expected. Note that similar magnetic loops were observed for the PFN--PZT and PFT--PZT ceramics. However, the remanent magnetization ascribed to ferromagnetism was a few times bigger than that in our ceramics.

The results are valid for other multiferroics having superparamagnetic and ferroelectric phases. The study performed by us demonstrated that multiferroics having superparamagnetic and ferroelectric phases can be considered as promising materials for various applications along with composite multiphase (ferroelectric/ferromagnetic) layered structures. Since the ME response in multiferroics having superparamagnetic phase is proportional to $\rd M^2/\rd H$, its value can be amplified by many orders due to a sharp change of magnetization with the field. The discovery of physical mechanism responsible for a giant ME response opens the way for obtaining novel multiferroics, both single-phase and composite, necessary for modern electronic technological devices. In particular, superparamagnetic magnetization obtained for a composite based on carbon powder with nanopores filled with Ni might probably be obtained using some ferroelectric material (e.g., KNbO$_3$ with large region of FE phase) instead of carbon. On the other hand, in single-phase multiferroic on the base of nanosized particles of Fe$_3$O$_4$ having a superparamagnetic phase, the ferroelectric phase can be induced by annealing in the nitrogen atmosphere, producing oxygen vacancies, which due to the Vegard effect is capable of introducing the FE phase.

\section{Conclusion}

We reviewed the recent trends of nanoferroics and multiferroics studies. The main attention is paid to spontaneous flexoeffects and reentrant phase in nanoferroics and to recently discovered giant magnetoelectric effect in multiferroics. In particular, ME coefficient measured by the authors of references~\cite{gli20,lag20} at room temperature is equal to $\beta = 0.54 \cdot 10^{-15}$ s/\AA~for (PFT)$_x$(PMN$_{0.7}$--PT$_{0.3}$)$_{1x}$ while its value for PFN--PT appeared to be much smaller, namely $\beta \sim 10^{-18}$ s/\AA~and its value for BiFeO$_3$ is $\beta = 2.54 \cdot 10^{-19}$~s/\AA. We have to note that both these phenomena (see references~\cite{mor21,gli08}) as well as the appearance of morphotropic phase in a relaxor due to the influence of the oxygen vacancies (see~\cite{gli18}) were proposed by the author of this review for the first time in the world science. The same statement is also correct for the recently discovered giant magnetoelectric effect in multiferroics (see~\cite{gli20}).


\bibliographystyle{cmpj}

%
%

\ukrainianpart

\title{Нові тенденції у нанофізиці фероїків, релаксорів та мультифероїків}
\author{М. Д. Глинчук, Л. П. Юрченко, Є. А. Єлісєєв}
\address{Інститут проблем матеріалознавства НАН України, вул. Омеляна Пріцака 3, 03142 Київ, Україна}

\makeukrtitle

\begin{abstract}
\tolerance=3000%
Огляд охоплює теоретичні та експериментальні результати, отримані за останні роки авторами за допомогою комплексного дослідження нанофероїків, релаксорів та мультифероїків. Основна увага приділена спонтанному флексоелектричному ефекту і реентрант-фазі в нанофероїках, а також нещодавно відкритому гігантському магнітоелектричному ефекту в мультифероїках.
\keywords фероїки, мультифероїки, фазові переходи, магнітоелектрика

\end{abstract}

\end{document}